\pdfoutput=1
\documentclass[pra,aps,groupedaddress,showpacs,letterpaper,notitlepage,longbibliography,floatfix]{revtex4-1}

\usepackage{amsmath,amsfonts,amssymb}

\usepackage{graphicx}

\DeclareGraphicsExtensions{.pdf}

\usepackage{times}

\makeatletter

\begin{document}

\title{One-Dimensional Waveguide Coupled to Multiple Qubits:\\
Photon-Photon Correlations
}

\author{Yao-Lung L. Fang}
\author{Huaixiu Zheng}
\author{Harold U. Baranger}
\email{baranger@phy.duke.edu}

\affiliation{\textit{Department of Physics, Duke University, P.O. Box 90305, Durham, North Carolina 27708, USA}}

\date{\today}


\begin{abstract}
For a one-dimensional (1D) waveguide coupled to two or three qubits, we show that the photon-photon correlations have a wide variety of behavior, with structure that depends sensitively on the frequency and on the qubit-qubit separation $L$. We study the correlations by calculating the second-order correlation function $g_2(t)$ in which the interference among the photons multiply scattered from the qubits causes rich structure. In one case, for example, transmitted and reflected photons are both bunched initially, but then become strongly anti-bunched for a long time interval. We first calculate the correlation function $g_2(t)$ including non-Markovian effects and then show that a much simpler Markovian treatment, which can be solved analytically, is accurate for small qubit separation. As a result, the non-classical properties of microwaves in a 1D waveguide coupled to many superconducting qubits with experimentally accessible separation $L$ could be readily explored with our approach.
\end{abstract}

\maketitle

\begin{figure*}[b]
	\centering
		\includegraphics[width=1.0\textwidth]{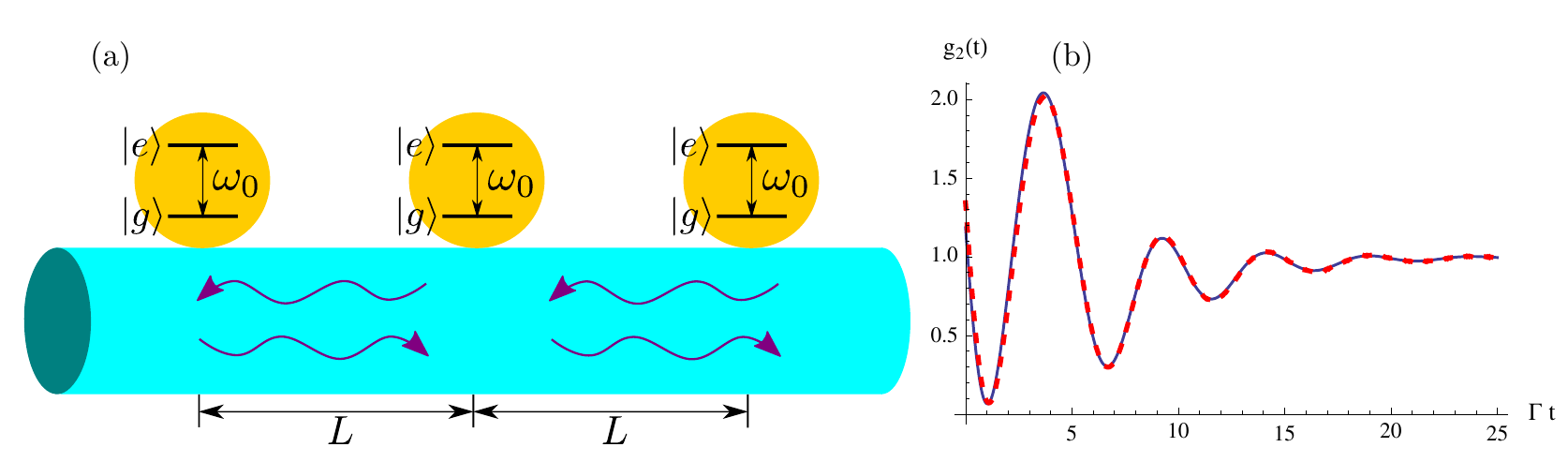}
\caption{\raggedright Quantum beats in three qubit system. (a) Schematic diagram of the 1D waveguide system coupled to 3 identical qubits with separation $L$. (b)~Comparison between the Markovian approximation (solid blue curve) and full numerical results (red dashed line) for $g_2(t)$ of reflected photons with $N\!=\!3$, $k_0L\!=\!\pi/2$, $T\!=\!50\%$, and $\Gamma'\!=\!0$.}
\label{fig:SchematicComparison}
\end{figure*}

\section{Introduction}
One-dimensional (1D) waveguide-QED systems are currently generating increasing interest---systems in which photons confined in one-dimension interact with one or several two-level systems (qubits). Part of the motivation comes from the striking quantum optics effects that can be seen in these strongly coupled systems \cite{YudsonJETP84,YudsonJETP85,ChangPRL06,ShenPRL07,ShenPRA07,YudsonPRA08,
WitthautNJP10,ZhengPRA10,HouNoriPRA10,RephaeliPRA11,RoyPRL11,RoyPRA11,
ShiSunPRA11,ZhengPRL11,ZhengPRA12,RephaeliPRL12,ZhengPRL13,MoeferdtOL13,
LalumierePRA13}.
Another motivating factor is the promise of waveguide-QED systems for quantum information processing 
\cite{ShenPRL05,ChangNatPhy07,ZhouPRL08,LongoPRL10,KolchinPRL11,EichlerWallraffPRA12,ZhengOL13,ZhengPRL13a}. 
Finally, a key driver of the interest in waveguide-QED systems is the tremendous experimental progress that has been made recently in a number of systems \cite{AkimovNat07, BajcsyPRL09, BabinecNatNanotech10, ClaudonNatPhoton10,BleusePRL11,LauchtPRX12,
AstafievSci10, AstafievPRL10,EichlerPRL11,HoiPRL11,HoiPRL12,EichlerPRL12,HoiPRL13,vanLooScience13}. 
Perhaps the leading system for waveguide-QED investigations and applications is an open microwave transmission line coupled to superconducting qubits \cite{AstafievSci10,AstafievPRL10,WallraffNat04,SchoelkopfNat08,EichlerPRL11,HoiPRL11}. While much of the work to date has focused on systems in which there is a single qubit, 
and there is a growing literature on the case of two qubits \cite{DzsotjanPRB10,TudelaPRL11,
DzsotjanPRB11,vanLooScience13,RephaeliPRA11,GonzalezTudelaPRL13,ZhengPRL13,GonzalezBallestroNJP13,RoySciRep13,LalumierePRA13}, an important future direction for both fundamental effects and possible applications is to study a waveguide coupled to multiple (or many) qubits. As a step in this direction, here we compare and contrast results for one, two, and three qubits coupled to a waveguide [see Fig.\,\ref{fig:SchematicComparison}(a)], focusing in particular on the generation of photon-photon correlations. 

Correlations between photons are a key signature of non-classical light. They are often characterized by the second-order correlation function (photon-photon correlation function) $g_2(t)$ where $t$ is the observation time between the two photons (see below for precise definition) \cite{LoudonQTL03}. The uncorrelated, classical value is $g_2\!=\!1$ (obtained, for example, for a coherent state). Bunching of photons, $g_2\!>\!1$, often occurs due to the bosonic nature of photons but anti-bunching, $g_2\!<\!1$, also occurs \cite{LoudonQTL03}. In recent experiments, $g_2(t)$ of microwave photons coupled to superconducting qubits was measured, and both bunching and anti-bunching were observed \cite{LangPRL11,HoiPRL12}. In a multi-qubit situation, one expects to have interference between the various scattered partial waves; interference effects in the photon-photon correlations $g_2(t)$ are known as ``quantum beats'' \cite{FicekPRA90}.

In this paper, we first present our method of calculation, which exploits a bosonic representation of the qubits in the rotating wave approximation. We obtain a complicated yet analytic result for $g_2 (t)$ in the Markovian limit and show, by comparison with the full numerical result, that it is adequate for small, experimentally accessible separations between the qubits. In presenting results, we focus on an off-resonant case in which single photons have equal probability of being transmitted or reflected, and take the separation between qubits, denoted $L$, to be either $\lambda_0/4$ or $\lambda_0/8$ where $\lambda_0$ is the wavelength of a photon at the qubit resonant frequency. We find several striking features in $g_2 (t)$: First, for $L\!=\!\lambda_0/4$, the transmitted photons are largely bunched for all times and become more strongly bunched as the number of qubits increases, while the reflected photons oscillate between strong bunching and anti-bunching, showing particularly strong quantum beats in the three qubit case. Second, for $L\!=\!\lambda_0/8$, we find the surprising situation that both transmitted and reflected photons are bunched at $t\!=\!0$ but then become anti-bunched for a large time interval. This suggests that the photons in this case become organized into bursts.

\section{Method}
The Hamiltonian describing $N$ identical qubits coupled to a 1D waveguide [see Fig.~\ref{fig:SchematicComparison}(a)] is, in the rotating wave approximation,
\begin{align}
H_0&=\hbar(\omega_0-i\Gamma^{\prime}/2)\sum_{i=1}^N\sigma^+_i\sigma^-_i
-i\hbar c\int dx \left[a^\dagger_\text{R}(x)\frac{d}{dx}a_\text{R}(x)-a^\dagger_\text{L}(x)\frac{d}{dx}a_\text{L}(x)\right]\nonumber\\
&\,\,+\sum_{i=1}^N\sum_{\alpha=\text{L,R}} \hbar V \int dx\; \delta(x-l_i)\left[a^\dagger_\alpha(x)\sigma^-_i+a_\alpha(x)\sigma^+_i\right],
\end{align}
where $\sigma^\pm_i$ are the raising/lowering operators for $i$-th qubit, $l_i$ is its position which is fixed by $L\!=\!l_{i+1}-l_{i}$, $\omega_0$ is the transition frequency of the qubit, and $\Gamma'$ is the decay rate to channels other than the waveguide. 
The spontaneous decay rate to the waveguide continuum is given by $\Gamma\!=\!2V^2/c$. In the waveguide QED context, ``strong coupling'' signifies that the spontaneous decay rate to the waveguide is much faster than the decay to all other modes, namely that the Purcell factor is large, $P\!\equiv\!\Gamma/\Gamma' \gg 1$.

To find $g_2(t)$, we first obtain the two-photon eigenstate of $H_0$. As discussed in Ref.~\cite{ZhengPRL13}, it is convenient to use a bosonic representation of the qubits that includes an on-site interaction, 
\begin{equation}
H=H_0 + V,\quad V=\frac{U}{2}\sum_{i=1}^N d^\dagger_i d_i(d^\dagger_i d_i-1)
\label{eq:onsite_interaction} \;.
\end{equation}
The raising/lowering operators $\sigma^\pm_i$ in $H_0$ are replaced by the bosonic creation/annihilation operators $d^\dagger_i$ and $d_i$, respectively. One then takes $U\!\to\!\infty$ in the end to project out occupations greater than $1$. 
In this bosonic representation, the $U\!=\!0$ case corresponds to a non-interacting Hamiltonian and can readily be solved. In terms of the non-interacting wavefunctions and Green functions, a formal expression for the two-photon ``interacting'' wavefunction in the $U\!\to\!\infty$ limit can be obtained; this then is the solution to the waveguide QED problem in which we are interested. 
Finally, the two-photon wavefunction together with the one-photon wavefunction yields $g_2(t)$ for a weak incident coherent state.
More details of this procedure are given in the appendices. 

The Markovian approximation allows a considerable simplification of the final result \cite{ZhengPRL13}. In the present context, the Markovian approximation consists of an approximate treatment of certain interference terms valid for small separation between the qubits. In the formal expression for the two-photon wavefunction discussed above, there is an integral over the non-interacting wavefunctions which generally must be performed numerically. The non-interacting wavefunctions naturally involve interference factors $e^{i2kL}$ that make this integral difficult. However, if the qubits are close enough, $k$ may be replaced by $k_0 \!=\! \omega_0/c$, allowing the integral to be performed analytically using contour integration (the analytic expression of the final result is lengthy, so we just give the steps of the derivation in the appendices as well as the $N\!=\!2$ result as an example). All of the results in this paper are obtained in the regime where this is valid. An example of the checks we have made is shown in Fig.~\ref{fig:SchematicComparison}(b): the full numerical result is in good agreement with that from the small separation approximation.

We compare the one, two, and three qubit cases: $N\!=\!1$, $2$, or $3$. In order to make a fair comparison, the typical transmission through the system in the three cases should be the same; otherwise, the lower probability of finding a photon in one case compared to another will affect $g_2$. We therefore consider off-resonance cases (i.e.\ $\omega\!\neq\!\omega_0$ where $\omega$ is the incoming photon frequency) in which the single-photon transmission probability $T$ is fixed. Because the single-photon transmission spectrum depends on the number of qubits, the frequency used is different in the three cases $N\!=\!1$\textendash$3$. Due to the asymmetry of the single-photon transmission spectrum in certain cases, the criterion used throughout this work is to pick up the frequency closest to $\omega_0$ so that $g_2(0)$ is the largest.

\begin{figure*}
	\centering
		\includegraphics[width=1.0\textwidth]{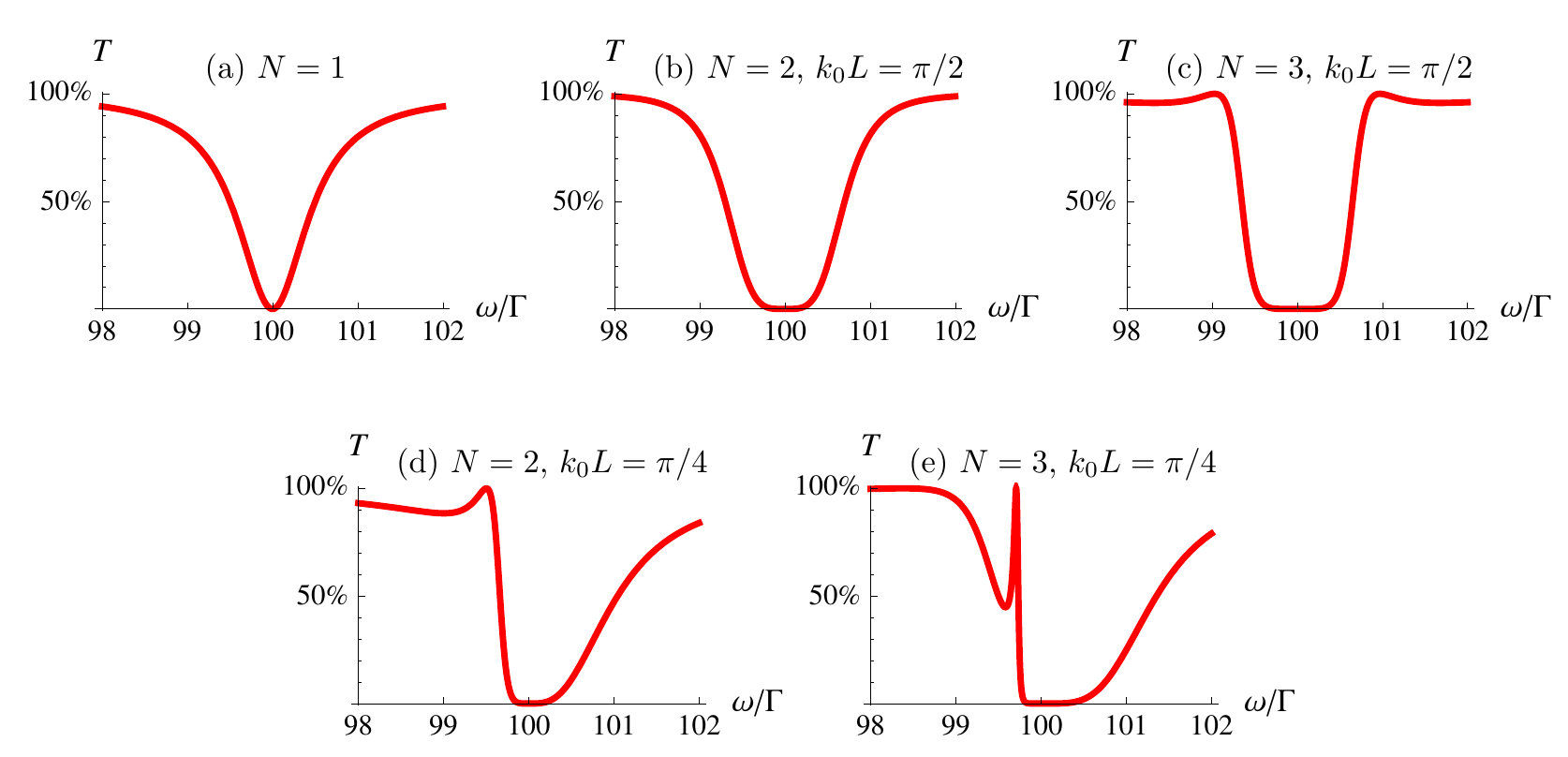}
\caption{Transmission spectra near the qubit resonant frequency ($\omega_0 \!=\!100\,\Gamma$) for an incident single-photon Fock state in the five situations studied here.}
\label{fig:transmission-spectrum}
\end{figure*}

In the following results, we consider $N\!=\!1$, $2$, or $3$; $k_0L\!=\! \pi/4$ or $\pi/2$; and $T\!=\!50\%$. The single photon transmission curves used to choose the photon frequency $\omega$ are shown in Fig.\,\ref{fig:transmission-spectrum}. We use $\Gamma$ as our unit of frequency, take $\omega_0\!=\!100\Gamma$, and consider the lossless case, $\Gamma^{\prime}\!=\!0$.

\section{Results}

\begin{figure*}
	\centering
		\includegraphics[width=1.0\textwidth]{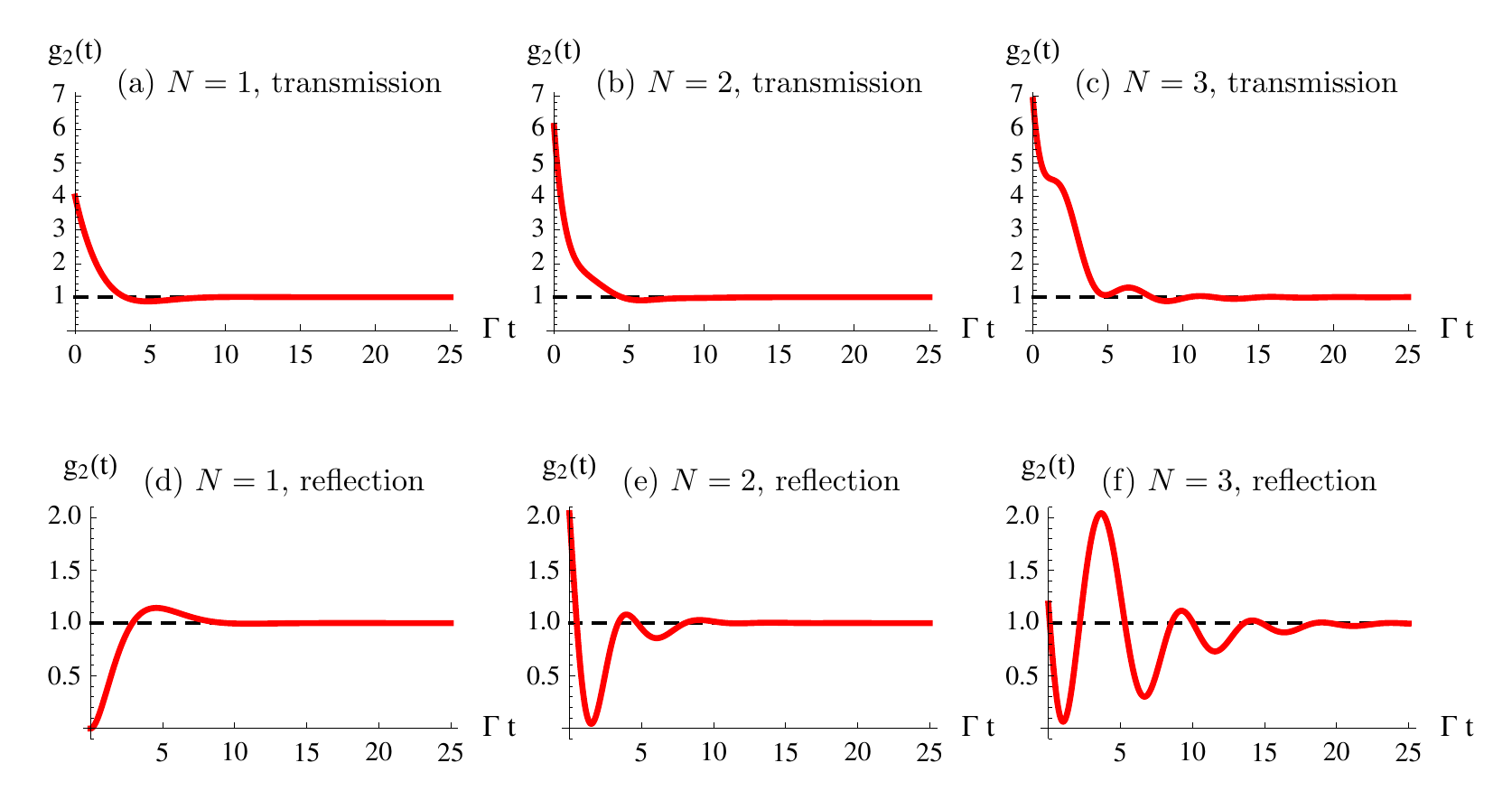}
\caption{Second-order correlation function, $g_2(t)$, calculated with a weak incident coherent state for spacing $k_0L\!=\!\pi/2$. First row is for transmitted photons, second row for reflected photons. The columns correspond to $N\!=\!1$, $2$, or $3$ qubits coupled to the waveguide. The photon frequency is chosen so that $T\!=\!50\%$. The result for uncorrelated photons, $g_2\!=\!1$, is marked (dashed line) for comparison. In the three qubit case, note the strong bunching in transmission [panel (c)] and striking quantum beats in reflection [panel (f)]. 
}
\label{fig:Piover2-T50}
\end{figure*}

\begin{figure*}
	\centering
		\includegraphics[width=1.0\textwidth]{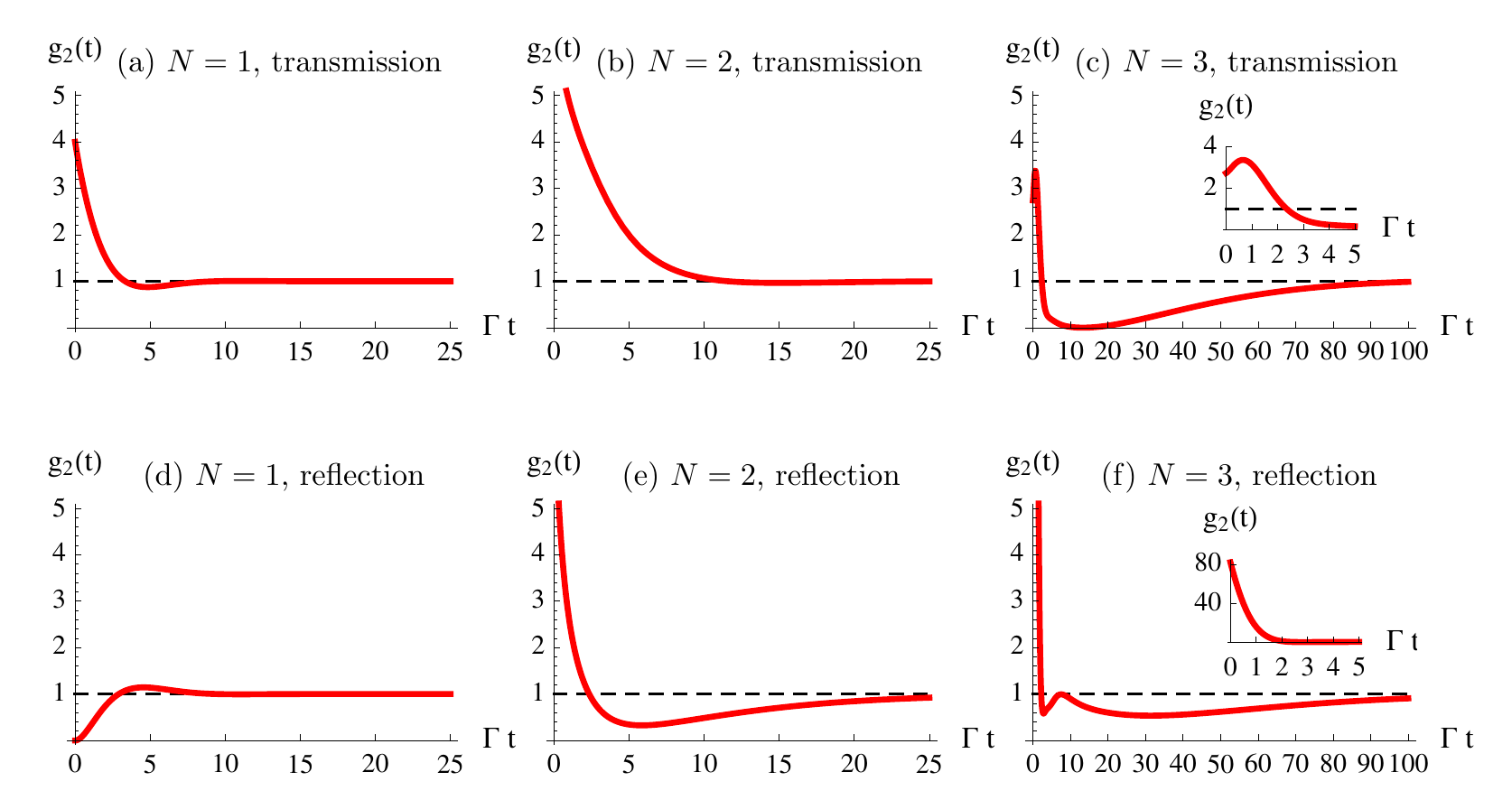}
\caption{Second-order correlation function, $g_2(t)$, calculated with a weak incident coherent state for spacing $k_0L\!=\!\pi/4$. First row is for transmitted photons, second row for reflected photons. The columns correspond to $N\!=\!1$, $2$, or $3$ qubits coupled to the waveguide. The photon frequency is chosen so that $T\!=\!50\%$. The result for uncorrelated photons, $g_2\!=\!1$, is marked (dashed line) for comparison. In the three qubit case, note the strong bunching in reflection [panel (f)] and long anti-bunching interval after the initial bunching in transmission [panel (c)].
}
\label{fig:Piover4-T50}
\end{figure*}

The results for a single qubit, shown in Fig.\,\ref{fig:Piover2-T50} panels (a) and (d), provide a point of comparison for the two and three qubit cases discussed 
below; throughout we consider the response to an incident weak coherent state. 
Non-classical light in a waveguide produced by a single qubit has been extensively investigated theoretically \cite{YudsonJETP84,YudsonJETP85,ShenPRL07,ShenPRA07,YudsonPRA08,ZhengPRA10,RoyPRL11,RoyPRA11,ShiSunPRA11} as well as experimentally with microwave photons \cite{HoiPRL12}. We see that for our chosen detuning such that $T\!=\!50\%$, the transmitted field shows bunching while the reflected field is anti-bunched. The correlation decays to its classical value (namely, $1$) quickly and with little structure. For this reason the single value $g_2(0)$ is a good indication of the nature of the correlations overall. Note that in panel (d), $g_2(0)\!=\!0$ due to the inability of a single excited qubit to release two photons at the same time.

For $N\!=\!2$ or $3$, we start by considering the case $k_0L\!=\!\pi/2$, in which case the qubits are separated by $\lambda_0/4$; the results are shown in Fig.\,\ref{fig:Piover2-T50}. The presence of quantum beats coming from interference among the partial waves scattered by the qubits is clear, especially for three qubits. In the transmitted wave, photon bunching is considerably enhanced in magnitude and extends for a longer time (compared to a single qubit). In reflection, $g_2(t)$ develops a striking oscillation between strongly bunched and anti-bunched [panel (f)]. Such behavior in $g_2$ suggests that the photons become organized periodically in time and space. 

Turning now to the case $k_0L\!=\!\pi/4$, we see in Fig.\,\ref{fig:Piover4-T50} that the behavior is completely different. First, the quantum beats largely disappear in both transmission and reflection. Instead, for $N\!=\!3$ we see that both the reflected and transmitted photons are initially bunched, in the reflected case quite strongly bunched. The initial bunching is followed in both cases by anti-bunching. This anti-bunching is dramatic for the transmitted photons: strong anti-bunching persists for a time interval of several tens of $\Gamma^{-1}$ (the natural unit of time in our problem). Initial bunching followed by a long interval of anti-bunching suggests that the photons are organized into bursts. 

The different behavior for $k_0L\!=\!\pi/4$ compared to $k_0L\!=\!\pi/2$ can be traced to a difference in the structure of the poles of the single photon Green function (see, e.g., the discussion in Ref.~\cite{ZhengPRL13}). For instance in the $N\!=\!2$ cases, for $k_0L\!=\!\pi/2$ there are two dominant poles that have the same decay rate but different real frequencies, leading to maximum interference effects between those two processes. In contrast, for $k_0L\!=\!\pi/4$, the poles have very different decay rates; the one decaying most rapidly yields the sharp initial bunching, while the one with the slowest decay produces the long time anti-bunching.

\begin{figure*}
		\centering
                \includegraphics[width=1.0\textwidth]{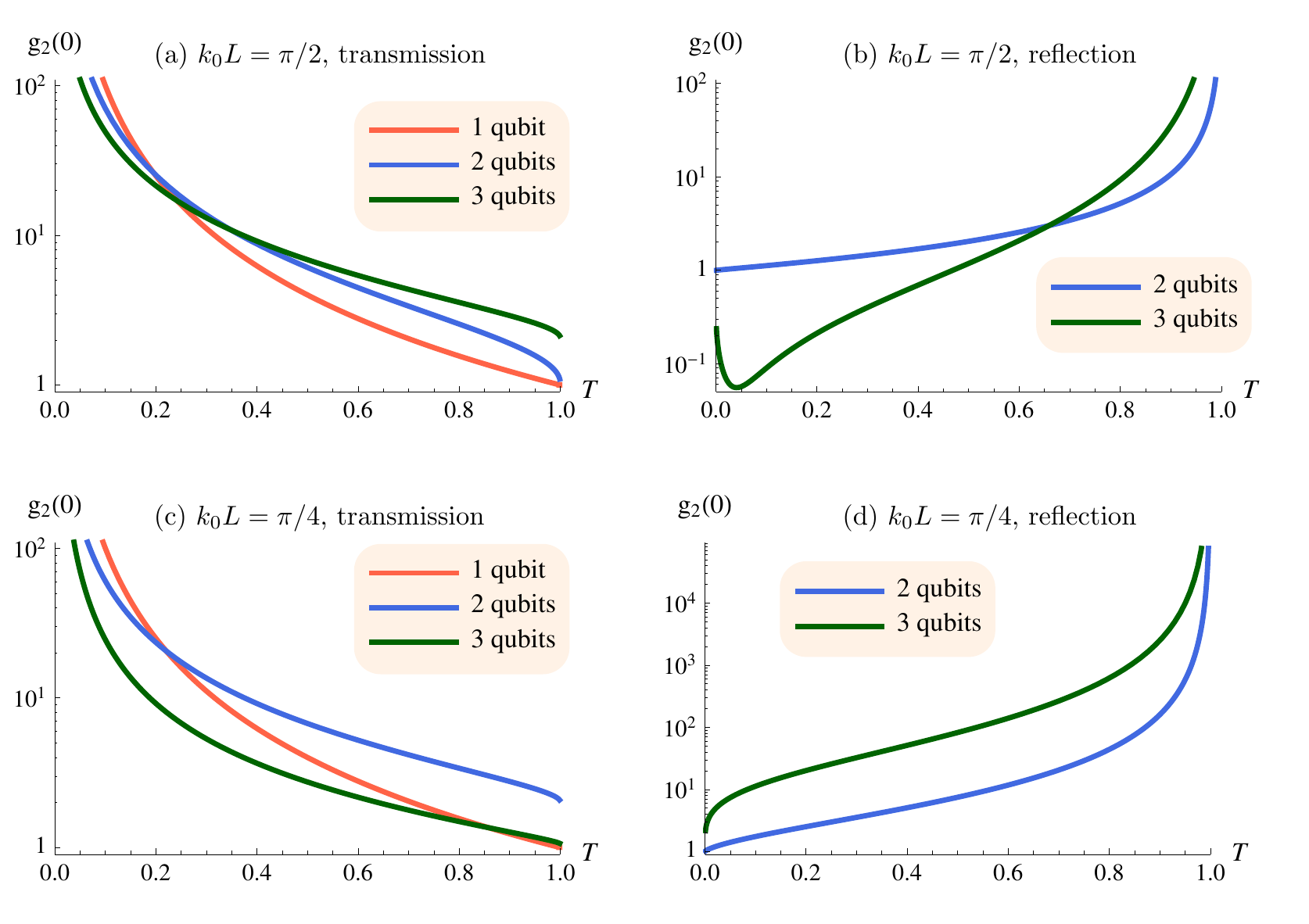}
        \caption{
The initial second-order correlation, $g_2(0)$ (on a logarithmic scale), calculated with a weak incident coherent state as a function of the single-photon transmission probability, $T$, for different number of qubits. The first (second) row is for $k_0L\!=\!\pi/2$ ($\pi/4$); the first (second) column is for transmitted (reflected) photons. For reflected photons with $N\!=\!1$, $g_2(0)\!=\!0$ for all $T$ and hence is not plotted. For a wide range of parameters, both transmitted and reflected photons are bunched.
}
        \label{fig:g2(0)}
\end{figure*}

\begin{figure*}
		\centering
                \includegraphics[width=1.0\textwidth]{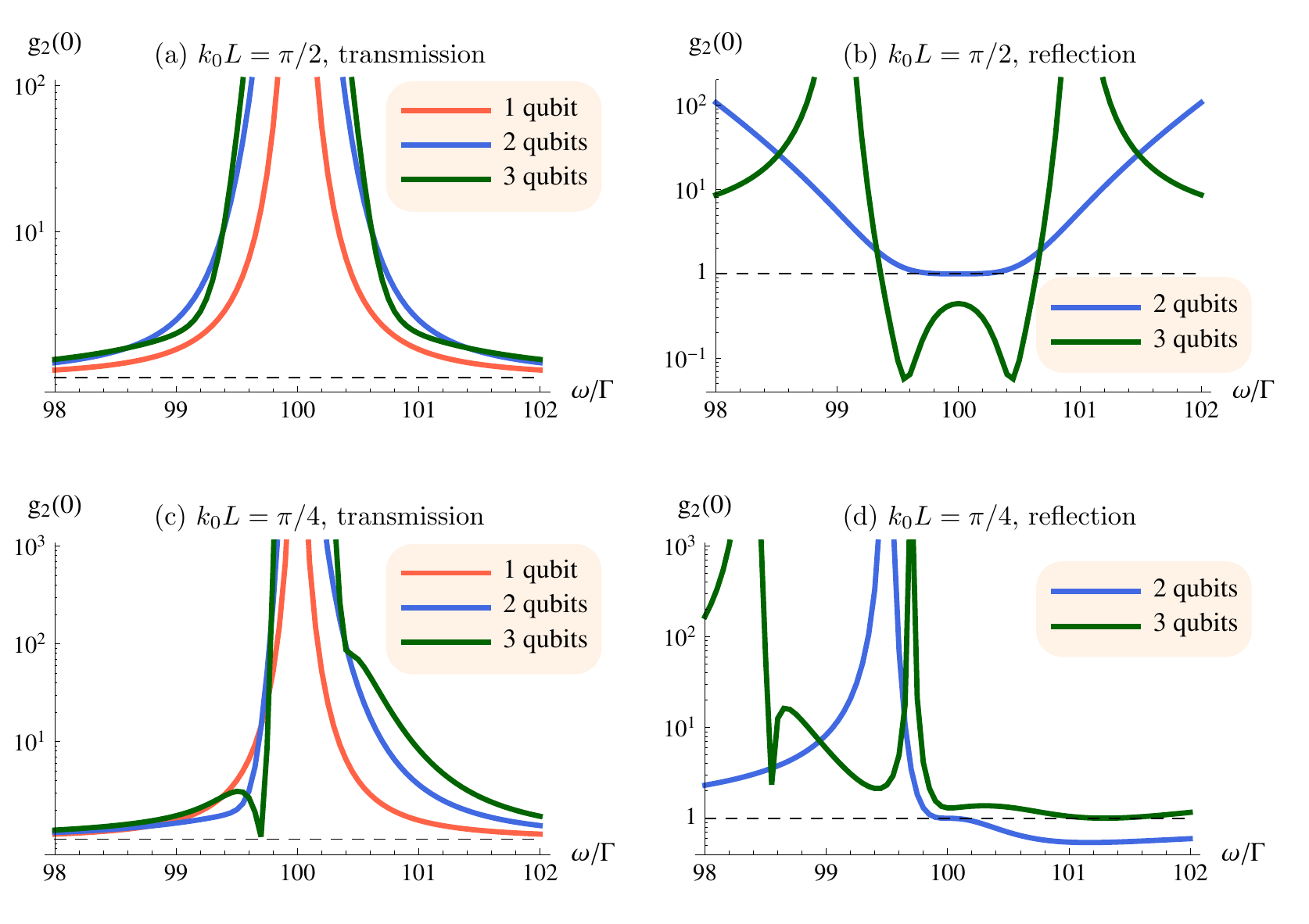}
        \caption{The initial second-order correlation, $g_2(0)$ (on a logarithmic scale), calculated with a weak incident coherent state as a function of the frequency $\omega$ near the qubit resonant frequency ($\omega_0 \!=\!100\,\Gamma$) for different numbers of qubits. The first (second) row is for $k_0L\!=\!\pi/2$ ($\pi/4$); the first (second) column is for transmitted (reflected) photons. The black, dashed line indicates the classical value (i.e., $g_2=1$). For reflected photons with $N\!=\!1$, $g_2(0)\!=\!0$ for all $\omega$ and hence is not plotted. Note that as for the single-photon transmission in Fig.~\ref{fig:transmission-spectrum}, $g_2(0)$ is symmetric about $\omega_0$ for $k_0L=\pi/2$ but asymmetric in the $k_0L=\pi/4$ case. For a wide range of parameters, both transmitted and reflected photons are bunched.}
        \label{fig:g2(0) vs frequency}
\end{figure*}

To study how the correlations depend on the frequency of the photons, we show the initial correlation, $g_2(0)$, in Fig.\,\ref{fig:g2(0)} and \ref{fig:g2(0) vs frequency}. Because of the oscillating structure in $g_2(t)$ when there are multiple qubits, $g_2(0)$ is not necessarily a good indication of the behavior at later times; nevertheless, the degree of initial bunching or anti-bunching is a physically important and measurable quantity. For a fair comparison between the $N\!=\!1$, $2$, and $3$ cases, we first plot $g_2(0)$ as a function of the single photon transmission, $T$; see Fig.~\ref{fig:g2(0)}. To match the desired $T$ with an off-resonant photon frequency we follow the following procedure: Starting near the resonant frequency $\omega_0$ (where $T\!=\!0.1\%$), we scan toward smaller frequencies until $T\!=\!99.9\%$ is reached. We then use frequencies within the scanned range to calculate $g_2(0)$ for both transmission and reflection as a function of $T$ at $k_0L\!=\!\pi/2$ and $\pi/4$. 
Another way of presenting the data is to simply plot $g_2(0)$ directly as a function of frequency, as in Fig.~\ref{fig:g2(0) vs frequency}. By comparing with Fig.~\ref{fig:transmission-spectrum}, we see that the method above for selecting the range of frequencies to use in making Fig.~\ref{fig:g2(0)} selects the range with largest $g_2(0)$ for a given value of $T$. Finally, note that for reflection from one qubit, $g_2(0)\!=\!0$ in all cases, as mentioned above, and so is not plotted in panels (b) and (d) of both figures. 

Several general trends are clear from Fig.\,\ref{fig:g2(0)}. Bunching is favored over anti-bunching for both $N\!=\!2$ and $3$. As the single photon transmission increases, $g_2(0)$ decreases for transmission but generally increases for reflection. Opposite trends for transmission and reflection are natural based on the simple argument that incoming uncorrelated photons divide between transmitted and reflected ones so that bunching in one implies anti-bunching in the other. Clearly, this simple argument does not apply here; indeed, it is striking and surprising that for a broad range of parameters \textit{both transmitted and reflected photons are bunched.} 

Trends as the number of qubits increases from $1$ to $3$ are also evident in Fig.~\ref{fig:g2(0)}. In panels (a) and (d) the trend is monotonic: For $k_0L\!=\!\pi/4$, the reflected photons become tremendously bunched [panel (d)], whereas for $k_0L\!=\!\pi/2$ and transmitted photons [panel (a)], the curves cross at the same point indicating that the trend changes sign---increasing bunching as $N$ increases for $T\geq 0.25$ but decreasing bunching for smaller $T$. In the other two cases, panels (b) and (c), the trend as $N$ increases from $1$ to $3$ is not monotonic. For $k_0L\!=\!\pi/2$ and $T\leq 0.65$, the reflected photons switch from being anti-bunched to bunched to anti-bunched as $N$ changes from $1$ to $3$, but show increasing bunching for larger $T$ [panel (b)]. Finally, in panel (c) [$k_0L\!=\!\pi/4$ and transmitted photons], there is a monotonic trend toward less bunching for $T\leq0.25$ but non-monotonic behavior for larger transmission. 

From the explicit dependence on frequency shown in Fig.~\ref{fig:g2(0) vs frequency}, we see that bunching is generally favored even outside the frequency range chosen in Fig.~\ref{fig:g2(0)} (which in the $k_0L\!=\!\pi/4$ case is quite small ($<\Gamma$)). Comparing to the single photon transmission spectrum (see Fig.~\ref{fig:transmission-spectrum}), we point out two features: First, in the $k_0L\!=\!\pi/4$ case, $g_2(0)$ shows the asymmetry with respect to $\omega_0$ [panel (c) and (d)] seen in $T(\omega)$; this again can be traced to the asymmetric pole structure of the Green functions mentioned above. $g_2(0)$ is larger (for reflection) and varies more rapidly on the red-detuned side ($\omega\!<\!\omega_0$), which explains why we chose the frequency range use in Fig.~\ref{fig:g2(0)}. In fact, on the blue-detuned side ($\omega\!>\!\omega_0$) the structure in $g_2(t)$ is less dramatic, and it returns to $1$ faster (data not shown). Second, as also shown in Fig.~\ref{fig:g2(0)}, the peaks of $g_2(0)$ for transmission are locatd where $T=0$, while the peaks of reflected $g_2(0)$ are located where $T=1$. Note that the leftmost peak of $N\!=\!3$ in Fig.~\ref{fig:g2(0) vs frequency}(d) is completely due to the small denominator ($R=1-T\approx0$) at that point.

\section{Conclusion}
In this work, we have calculated the second-order correlation function, $g_2(t)$, for photons in a one-dimensional waveguide interacting with one, two or three qubits. By taking the separation between the qubits small, we are able to make a Markovian approximation which then allows an analytic solution. The small separation and small $N$ on which we focus means that these systems are within the range of current experimental capability \cite{LalumierePRA13}. 

The interference among the partial waves scattered from the qubits leads to a variety of behavior in $g_2$ that is sensitive to both the separation between the qubits ($L$) and the frequency of the incoming photons. As examples of the rich variety accessible in these waveguide QED structures, we mention three here in conclusion: (i)~For a wide range of parameters, both transmitted and reflected photons are initially bunched. (ii)~For reflected photons with $N\!=\!3$ and $k_0L\!=\!\pi/2$, $g_2(t)$ oscillates between bunching and anti-bunching [Fig.\,\ref{fig:Piover2-T50}(f)]. (iii)~For transmitted photons with $N\!=\!3$ and $k_0L\!=\!\pi/4$, initial strong bunching is followed by a long (i.e.\ $\sim\!30\,\Gamma^{-1}$) interval of antibunching [Fig.\,\ref{fig:Piover4-T50}(c)]. These last two observations suggest that some nascent organization of the photons may be occurring, providing an interesting direction for future research.

\begin{acknowledgments}
This work was supported by US NSF\,Grant\,No.~PHY-10-68698. H.Z.\ is supported by a John T.\ Chambers Fellowship from the Fitzpatrick Institute for Photonics at Duke University.
\end{acknowledgments}

\appendix

\section{Two-photon interacting scattering eigenstate}
The single photon eigenstate $|\phi_1(k)\rangle_\alpha$ with $\alpha=\text{L,\,\,R}$ is by definition the eigenstate of $H_0$, i.e., $H_0|\phi_1(k)\rangle_\alpha=\hbar c k |\phi_1(k)\rangle_\alpha$, where
\begin{align}
|\phi_1(k)\rangle_\alpha &= \left[\int dx\left(\phi_\text{R}^\alpha(k,x)
a_\text{R}^\dagger(x)+\phi_\text{L}^\alpha(k,x)
a_\text{L}^\dagger(x)\right)+\sum_{i=1}^N e_i^\alpha(k)\sigma^+_i\right]|0\rangle,\\
\phi_\text{R}^\text{R}(k,x) &= \frac{e^{ikx}}{\sqrt{2\pi}}\left(\theta(l_1-x)+\sum_{i=1}^{N-1} t_i(k)\theta(x-l_i)\theta(l_{i+1}-x) 
+ t_N(k)\theta(x-l_N)\right)\!,\\
\phi_\text{L}^\text{R}(k,x) &= \frac{e^{-ikx}}{\sqrt{2\pi}}\left(r_1(k)\theta(l_1-x)+\sum_{i=2}^{N} r_i(k)\theta(x-l_{i-1})\theta(l_{i}-x)\right),\\
\phi_\text{R}^\text{L}(k,x) &= \frac{e^{ikx}}{\sqrt{2\pi}}\left( r_1(k)\theta(l_N-x)+\sum_{i=2}^{N} r_i(k)\theta(x-l_{N-i+1})\theta(l_{N-i+2}-x)\right),\\
\phi_\text{L}^\text{L}(k,x) &= \frac{e^{-ikx}}{\sqrt{2\pi}}\bigg(\theta(x-l_N) + \sum_{i=1}^{N-1} t_i(k)\theta(x-l_{N-i})\theta(l_{N-i+1}-x) 
+t_N(k)\theta(l_1-x)\bigg),
\end{align}
and the incoming photon travels in the $\alpha$-direction with wavevector $k$. The single photon transmission amplitude is given by $t_N(k)$ and the reflection amplitude by $r_1(k)$. Note that the positions of the qubits are chosen to be symmetric with respect to the origin, i.e., $l_{N-i+1}\!=\!-l_{i}$, in order to take advantage of parity symmetry. Setting $\hbar\!=c\!=1$ from now on, we have for $N\!=\!2$ \cite{ZhengPRL13}
\begin{align}
t_2(k)&=\frac{4 (k-\omega_0 )^2}{(i \Gamma +2 k-2 \omega_0 )^2+\Gamma ^2 e^{2 i k L}}\\[1mm]
r_1(k)&=\frac{\Gamma  \left(\Gamma -e^{2 i k L} (\Gamma +2 i k-2 i \omega_0 )-2 i k+2 i \omega_0 \right)}{\Gamma ^2 e^{3 i k L}+e^{i k L} (i \Gamma +2 k-2 \omega_0 )^2}\\[1mm]
e_1^\text{R}(k)&=-\frac{i \sqrt{\Gamma } e^{-\frac{1}{2} i k L} \left(\Gamma  \left(-1+e^{2 i k L}\right)+2 i k-2 i \omega_0 \right)}{\sqrt{\pi } \left((i \Gamma +2 k-2 \omega_0 )^2+\Gamma ^2 e^{2 i k L}\right)}\\[1mm]
e_2^\text{R}(k)&=\frac{2 \sqrt{\Gamma } e^{\frac{i k L}{2}} (k-\omega_0 )}{\sqrt{\pi } \left((i \Gamma +2 k-2 \omega_0 )^2+\Gamma ^2 e^{2 i k L}\right)}.
\end{align}
The corresponding result for $N\!=\!3$ is
\begin{align}
t_3(k)&=\frac{8 (k-\omega_0 )^3}{\eta_+^3+2 \Gamma^2 \eta_+ e^{2 i k L} +\Gamma^2 \eta_- e^{4 i k L} }\\
r_1(k)&=\frac{\Gamma \left(\eta_+^2 + 2 e^{2 i k L} \left(\Gamma ^2+2 (k-\omega_0 )^2\right) + \eta_-^2 e^{4 i k L} \right) }
{2i \Gamma ^2 \eta_+ e^{4 i k L} + i\Gamma ^2 \eta_- e^{6 i k L} + i \eta_+^3 e^{2 i k L} }\\
e_1^\text{R}(k)&=\frac{\sqrt{\Gamma } e^{-i k L} \left(i \eta_+^2 + 2 \Gamma  e^{2 i k L} (i \Gamma +k-\omega_0 ) + \Gamma \eta_- e^{4 i k L} \right)}
{\sqrt{\pi } \left(i \eta_+^3 + 2i \Gamma ^2 \eta_+ e^{2 i k L} + i \Gamma ^2 \eta_- e^{4 i k L} \right)}\\
e_2^\text{R}(k)&=\frac{2 \sqrt{\Gamma } (k-\omega_0 )}
{\sqrt{\pi } \left(\eta_+^2 + i \Gamma \eta_- e^{2 i k L} \right)}\\
e_3^\text{R}(k)&=\frac{4 \sqrt{\Gamma } e^{i k L} (k-\omega_0 )^2}
{\sqrt{\pi } \left(\eta_+^3 + 2 \Gamma ^2 \eta_+ e^{2 i k L} + \Gamma ^2 \eta_- e^{4 i k L} \right)}, 
\end{align}
with $\eta_\pm \equiv 2k - 2\omega_0 \pm i\Gamma$.
Note that we do not need the other amplitudes for the rest of this section. For $N\!=\!1$ results see, e.g., Ref.~\cite{ZhengPRA10}.

We can now construct the two-photon ``non-interacting'' eigenstate
\begin{equation}
|\phi_2(k_1, k_2)\rangle_{\alpha_1\alpha_2}=\frac{1}{\sqrt{2}}|\phi_1(k_1)\rangle_{\alpha_1}\otimes|\phi_1(k_2)\rangle_{\alpha_2}.
\end{equation}
As described in the Supplementary Material of Ref.~\cite{ZhengPRL13}, starting from the Lippmann-Schwinger equation
\begin{align}
|\psi_2(k_1, k_2)\rangle_{\alpha_1\alpha_2}=|\phi_2(k_1, k_2)\rangle_{\alpha_1\alpha_2}+G^R(E)V|\psi_2(k_1,k_2)\rangle_{\alpha_1\alpha_2},\\
G^R(E)=\frac{1}{E-H_0+i\epsilon},\qquad\qquad\qquad\qquad
\end{align}
where $V$ is given in Eq.~\eqref{eq:onsite_interaction} and $E$ is the two photon energy, one can derive the two-photon interacting eigenstate in the coordinate representation in the $U\!\rightarrow\!\infty$ limit:
\begin{align}
_{\alpha_1',\alpha_2'}\langle x_1, x_2 | \psi_2(k_1, k_2) \rangle_{\alpha_1,\alpha_2}
&=\,_{\alpha_1',\alpha_2'}\langle x_1, x_2 | \phi_2(k_1, k_2)\rangle_{\alpha_1,\alpha_2}\nonumber\\
&\quad-\sum_{i,j=1}^N
G^{\alpha_1'\alpha_2'}_i(x_1,x_2) \left(G^{-1}\right)_{ij}\langle d_j d_j|\phi_2(k_1, k_2)\rangle_{\alpha_1,\alpha_2}
\label{eq:two photon interacting eigenstate}
\end{align}
\begin{align}
G^{\alpha_1\alpha_2}_i(x_1,x_2)
&=\,_{\alpha_1,\alpha_2}\langle x_1, x_2 |G^R(E)|d_i d_i \rangle\nonumber\\
&=\sum_{\alpha_1',\alpha_2'}\int dk_1 dk_2\frac{_{\alpha_1,\alpha_2}\langle x_1, x_2|\phi_2(k_1,k_2)\rangle_{\alpha_1',\alpha_2'} \langle \phi_2(k_1,k_2)| d_i d_i\rangle}{E- (k_1+k_2)+i\epsilon}
\label{eq:GiRR}
\end{align}
\begin{equation}
G^{-1}=\begin{pmatrix}
  G_{11} & G_{12} & \cdots & G_{1N} \\
  G_{21} & G_{22} & \cdots & G_{2N} \\
  \vdots  & \vdots  & \ddots & \vdots  \\
  G_{N1} & G_{N2} & \cdots & G_{NN}
 \end{pmatrix}^{-1}
\end{equation}
\begin{equation}
G_{ij}=\langle d_i d_i|G^R(E)|d_j d_j\rangle
=\sum_{\alpha_1,\alpha_2}\int dk_1 dk_2\frac{\langle d_i d_i|\phi_2(k_1,k_2)\rangle_{\alpha_1,\alpha_2} \langle \phi_2(k_1,k_2)| d_j d_j\rangle}{E- (k_1+k_2)+i\epsilon}
\label{eq:Gij}.
\end{equation}
Note that $x_1$ and $x_2$ here refer to the positions of the photons.

By observing the structure of these Green functions, one would realize that given the following two pieces
\begin{align}
\langle d_i d_i|\phi_2(k_1,k_2)\rangle_{\alpha_1,\alpha_2}&=e_i^{\alpha_1}(k_1)e_i^{\alpha_2}(k_2)\\
_{\alpha_1',\alpha_2'}\langle x_1 x_2|\phi_2(k_1,k_2)\rangle_{\alpha_1,\alpha_2}&=\frac{1}{2}\left(\phi_{\alpha_1'}^{\alpha_1}(k_1,x_1)\phi_{\alpha_2'}^{\alpha_2}(k_2,x_2)+\phi_{\alpha_1'}^{\alpha_2}(k_2,x_1)\phi_{\alpha_2'}^{\alpha_1}(k_1,x_2)\right),
\end{align}
the whole prescription is complete and in principle one may numerically compute the two-photon interacting eigenstate Eq.~\eqref{eq:two photon interacting eigenstate} for any $N$.

Finally, to proceed with the Markovian approximation, we explicitly write down the integrands in Eqs.~\eqref{eq:GiRR} and \eqref{eq:Gij}, replace the factors $\exp(2ikL)$ by $\exp(2ik_0 L)$ therein, and do the double integral by standard contour integral techniques enclosing the poles in the upper half complex plane (for the $N\!=\!3$ case, for example, the denominator of each transmission amplitude is a cubic polynomial in $k$, so there are three roots). The $N\!=\!2$ case \cite{ZhengPRL13} could serve as an illustrative example owing to its relatively simple polynomial structure: For $k_0L=A\pi$ with $0\leq A \leq 1/2$, we have
\begin{align}
G_{11}&=\frac{e^{2 i \pi  A} \Gamma ^2+2 \eta^2}{2 \left(e^{2 i \pi  A} \Gamma ^2 \eta+\eta^3\right)}\\
G_{12}&=-\frac{e^{2 i \pi  A} \Gamma ^2}{2 \eta \left(\left(-1+e^{2 i \pi  A}\right) \Gamma ^2-4 i \Gamma  \omega_0 +E^2+2 i E (\Gamma +2 i \omega_0 )+4 \omega_0 ^2\right)}\\
G^\text{R,R}_1(x_1,x_2)&=-\Biggl\{ \Gamma  \Bigl[\beta_+ \left(-i \beta_- \Gamma +E-2 \omega_0 \right) \left(2 E-2\omega_0+i \gamma\right) \nonumber\\
&\quad +\beta_- e^{e^{i \pi  A} \Gamma  t} \left(-i \beta_- \Gamma +2 E-4 \omega_0 \right) \left(E+i \gamma\right)\Bigr] \nonumber\\
&\quad \times \exp \left(i E (t+x_1)-\frac{1}{2} \gamma t \right)\Biggr\}\biggl/8 \left(e^{2 i \pi  A} \Gamma ^2 \eta+ \eta^3\right)\\
G^\text{R,R}_2(x_1,x_2)&=i\Biggl\{\Gamma\Bigl[ \beta_+   \left(-i \beta_- 
\Gamma +E-2 \omega_0 \right) \left(\left(-2+e^{i \pi  A}\right) 
\beta_+ \Gamma +2 i E-4 i \omega_0 \right) \nonumber\\
&\quad+ \beta_-  e^{e^{i \pi  A} \Gamma  t} \left(-\left(e^{i \pi  A}+e^{2 i \pi  A}-2\right) \Gamma -2 i E+4 i \omega_0 \right) \left(E+i \gamma\right)\Bigr] \nonumber\\
&\quad \times\exp \left(-\frac{1}{2} \gamma t -i \pi  A+i E (t+x_1)\right)\Biggr\}\biggl/8 \left(e^{2 i \pi  A} \Gamma ^2 \eta+ \eta^3\right),
\end{align}
where $\eta\equiv E-2\omega_0+i\Gamma$, $\gamma \equiv (e^{i\pi A}+1)\Gamma+2i\omega_0$, and $\beta_\pm \equiv e^{i \pi  A} \pm 1$. During the two contour integrations, $x_1\!>\!l_2$ and $x_2\!=\!x_1+t$ (with $t\!>\!0$) are used. Due to parity symmetry, $G_{21}=G_{12}$, $G_{22}=G_{11}$, $G^\text{L,L}_1(-x_1,-x_2)=G^\text{R,R}_2(x_1,x_2)$ and $G^\text{L,L}_2(-x_1,-x_2)=G^\text{R,R}_1(x_1,x_2)$.

\section{Two-photon correlation function $g_2(t)$}
For a non-dispersive photonic field operator in the Heisenberg picture which satisfies $a^\dagger(x,t)=a^\dagger(x-ct)$, the two-photon correlation function $g_2(t)$ can be rewritten in the Schr\"{o}dinger picture as
\begin{align}
g_2(t)&=\frac{\langle\psi|a^\dagger_\alpha(x)a^\dagger_{\alpha'}(x+ct)a_{\alpha'}(x+ct)a_\alpha(x)|\psi\rangle}{\langle\psi|a^\dagger_\alpha(x)a_\alpha(x)|\psi\rangle
\langle\psi|a^\dagger_{\alpha'}(x+ct)a_{\alpha'}(x+ct)|\psi\rangle}\nonumber\\
&\approx\frac{\left|_{\alpha\alpha'}\langle x, x+ct| \psi_2(k_1,k_2)\rangle_\text{R,R}\right|^2}{\left|_{\alpha}\langle x| \phi_1(k_1)\rangle_{\text{R}}\right|^2 \left|_{\alpha'}\langle x+ct| \phi_1(k_2)\rangle_{\text{R}}\right|^2}\label{eq:definition of g2},
\end{align}
where $|\psi\rangle$ is the asymptotic output state and $\alpha=\alpha'=\text{R}$ for transmitted photons or $\alpha\!=\!\alpha'\!=\!\text{L}$ for reflected photons. The second equality holds if a weak incident coherent state (mean photon number $\bar{n}\ll 1$) with right-going photons is assumed---as is appropriate for comparison with an eventual experiment---such that we consider only two-photon states in the numerator and one-photon states in the denominator. The justification for the latter is twofold: (i)~In the numerator, the 0- and 1- photon states are eliminated by the annihilation operators, leaving the 2-photon sector untouched which, then, can be described by $_{\alpha,\alpha'}\langle x, x+ct|\psi_2(k_1, k_2)\rangle_\text{R,R}$. (ii)~In the denominator, the probability of having only one photon is much larger then having two, so that the factors $|_\text{R,R}\langle \psi_2(k_1, k_2)| a^\dagger_{\alpha}(x)a_{\alpha}(x)| \psi_2(k_1, k_2)\rangle_\text{R,R}|^2$ can be replaced by the single photon eigenstate $\left|_{\alpha}\langle x| \phi_1(k_1)\rangle_{\text{R}}\right|^2$ given that $k_1\!=\!k_2\!=\!E/2$ (i.e.~two identical incident photons). We are thus lead to an explicit expression for the photon-photon correlations in terms of the 1- and 2- photon states found using the method outlined in Appendix A.


%

\end{document}